\documentclass[runningheads]{llncs}

\usepackage{booktabs} % For formal tables

\usepackage{comment} 
\usepackage{subfigure}
\usepackage{booktabs}
\usepackage[utf8]{inputenc}
\usepackage[T1]{fontenc}
\usepackage{multirow}

\newcommand{\smalltilde}{\scriptsize$\sim$\normalsize}

\usepackage{amsmath} % usato per text{} dentro math mode
\usepackage{amssymb} % usato per insieme vuoto \varnothing
\usepackage[linesnumbered,lined,boxruled, nofillcomment]{algorithm2e} %,commentsnumbered,rightnl
\DontPrintSemicolon
\SetNlSkip{1.2em} % distanza tra numero di riga e testo algoritmo
\usepackage{graphicx}

\usepackage{url}

\begin{document}

\title{An Incremental Truth Inference Approach \\to Aggregate Crowdsourcing Contributions \\in Games with a Purpose}
\author{Irene Celino and Gloria {Re Calegari}}
\institute{Cefriel -- Politecnico di Milano \\
			Viale Sarca 226, 20126 Milano - Italy \\
			\small\texttt{\{irene.celino,gloria.re\}@cefriel.com}\normalsize}

\titlerunning{An Incremental Truth Inference to Aggregate Contributions in GWAPs}

\maketitle

\begin{abstract}
We introduce our approach for incremental truth inference over the contributions provided by players of Games with a Purpose: we motivate the need for such a method with the specificity of GWAP vs. traditional crowdsourcing; we explain and formalize the proposed process and we explain its positive consequences; finally, we illustrate the results of an experimental comparison with state-of-the-art approaches, performed on data collected through two different GWAPs, thus showing the properties of our proposed framework.
\end{abstract}

%%%%%%%%%%%%%%%%%%%%%%%%%%%%%%%%%%%%%%%%%%%%%%%%%%%%%%%%%%%%%%%%%%%%%%%%%%%%%%%%%%%%%%%%%%%%%%%%%%%%%%%%%%
\section{Introduction and related work}\label{sec:intro}

Truth inference algorithms~\cite{zheng2017truth} have been heavily explored in crowdsourcing to aggregate and make sense of workers' contributions. Most state-of-the-art algorithms (e.g., majority voting, expectation maximization~\cite{dawid1979maximum}, message passing~\cite{karger2011iterative}) are computed ex-post, i.e. all contributions are first collected and then aggregated, usually by means of iterative algorithms to infer the truth and estimate worker quality until convergence; this requires setting a-priori the number of repetitions of user labeling on each task, possibly collecting redundant information.

Variations of truth inference algorithm include scheduling approaches to optimize task assignment to workers, especially when micro-payment budget is an issue~\cite{karger2014budget,han2017budgeted}, and assessment of workers' skills  to improve answer quality, especially when tasks are very varied or have diverse levels of difficulty~\cite{allahbakhsh2013quality,difallah2013pick,yang2016modeling}. Related investigation exists on the evaluation of repeated labeling strategies~\cite{sheng2008get} to understand when it is more convenient to stop collecting user contributions; in that work, strong assumptions are made with respect to user accuracy and task difficulty, considered constant across examples; however, those premises do not always hold in practical settings. 

A Game with a Purpose or GWAP~\cite{von2008designing} is a well-known Human Computation approach~\cite{law2011human} to encourage users to execute tasks through entertainment. We can see GWAPs as a special crowdsourcing mechanisms in which \emph{workers} are actually \emph{players}, rewarded with fun instead of micro-payments by leveraging intrinsic motivation~\cite{ryan2000intrinsic}. Players are attracted and motivated by the game itself and often they are not even aware that their actions in the game play are exploited to produce the ``collateral effect'' of solving tasks.

Aggregating users' contribution is a key issue also in Human Computation systems like GWAPs. Originally, aggregation was based on simple agreement: in the ESP game~\cite{von2004labeling}, the very first GWAP ever released, players typed in textual labels to tag images and two agreeing users were enough to consider the label ``true''. Afterwards, ``ground truth'' tasks, i.e. problems with known solution, were introduced to check the quality of contributions to cope with random answers or malicious players~\cite{quinn2011human,ul2013effects}. 

In most crowdsourcing platforms (like Amazon Mechanical Turk\footnote{Cf. \url{https://www.mturk.com/}.} or Figure Eight\footnote{Cf. \url{https://www.figure-eight.com/}.}), tasks are assigned in batches or Human Intelligence Tasks (HITs) and workers are required to submit their answers within a specific time-frame in order to be eligible for payment~\cite{brabham2013crowdsourcing}. In contrast, in GWAPs contributions are collected as soon as a user decides to play the game: the flow of incoming answers is therefore subject to the ``appreciation'' of the game by players and a long-tail effect is very often recorded, with a few players playing a lot of rounds and the majority of participants being active for a few minutes only. Therefore, it is of utmost importance to exploit every single player's contribution and to infer truth in an incremental way, assigning the same task to the minimum sufficient number of different players.

The remainder is organized as follows: we give preliminaries and requirements for problem formulation in Section~\ref{sec:problem}; we describe our approach in Section~\ref{sec:approach} with the algorithm and its qualitative assessment; Section~\ref{sec:eval} presents a quantitative evaluation in comparison to baselines; Section~\ref{sec:concl} concludes the paper.

%%%%%%%%%%%%%%%%%%%%%%%%%%%%%%%%%%%%%%%%%%%%%%%%%%%%%%%%%%%%%%%%%%%%%%%%%%%%%%%%%%%%%%%%%%%%%%%%%%%%%%%%%%
\section{Problem formulation}\label{sec:problem}
In this section we formulate the problem, by giving some definitions and listing the requirements that the truth inference algorithm should fulfill. For simplicity of explanation, we specifically consider the case of multinomial classification tasks (with a pre-defined set of labels), but the approach can be easily extended to open labelling with no loss of generality.

% % % % % % % % % % % % % % % % % % % % % % % % % % % % % % % % % % % % 
\subsection{Definitions}
We consider a Game with a Purpose aimed to solve a set of tasks $T = \{ t_n \,|\, n=1,2,\ldots,N \}$. 
Each task is a labelling task, in which a label is assigned from a set of admissible values $V = \{ v_l \,|\, l=1,2,\ldots,L \}$. 

The GWAP is played by a set of users $U = \{ u_k \,|\, k=1,2,\ldots,K \}$. 
In each game round, a player is assigned a subset $T' \subset T$ of tasks to be solved. 
Given a set of ``ground truth'' tasks $G = \{ g_m \,|\, m=1,2,\ldots,M \}$  
for which the solution is known, in each game round the player is also given a set $G' \subset G$ of control tasks. 
The answers to control tasks are used to estimate the reliability $q_k$ of the player, which is useful to ``weight'' contributions on unsolved tasks during truth inference.

Player contributions are collected in a matrix $C = \{ c_{n,k} \,|\, n=1,2,\ldots,N \wedge k=1,2,\ldots,K\}$, 
initialized with null or zero values and filled with labels from $V$ whenever a player completes a task. The goal of the GWAP is not to completely fill up $C$, on the contrary $C$ should remain a sparse matrix, with the minimum possible number of players contributions (i.e., non-zero values) required to infer the ``true'' labels for the tasks.

Finally,  \emph{truth inference}  is a function applied on players' answers and reliability values to infer the result set $\hat{Y} = \{ \hat{y}_n \,|\, n=1,2,\ldots,N \}, \, \hat{y}_n \in V$ for each of the tasks in $T$. 
$\hat{Y}$ is computed by aggregation of users' contributions and is an estimate of the ``true'' unknown labelling $Y$ of the tasks.
Truth inference is \emph{incremental} if, at each new contribution from a GWAP player, a new estimation of $\hat{Y}$ is computed.

To understand if a task $t_i$ can be considered completed, truth inference computes a set of scores representing the confidence values on the association between $t_i$ and each possible labelling value $v_l$. In other words, the aggregation algorithm builds and updates a matrix of \emph{estimation scores} $S = \{ s_{n,l} \,|\, n=1,2,\ldots,N \wedge l=1,2,\ldots,L\}, \, s_{n,l} \in [0,1]$. % = \{ s_{n,l} \}_{N \times L}
As in record linkage literature~\cite{fellegi1969theory}, those scores start from $0$, and are incrementally increased according to user contributions. Each task $t_i$ is solved when the maximum of its scores $s_{i,*}$ (i.e. the $i$th row of matrix $S$) overcomes some threshold $\bar{s}$; the ``completion'' condition can be therefore  formulated as follows: 
\begin{equation}\label{eq:exit}
	 \forall t_i \in T \,\, \exists\, j \in \{1,2,\ldots,L\} \, | \, s_{i,j}=\,\texttt{max}(s_{i,*}) \, \wedge \, s_{i,j}>\bar{s} 
\end{equation}
Truth inference algorithms differ for their specific approach to update the matrix $S$ of scores when aggregating user contributions.

% % % % % % % % % % % % % % % % % % % % % % % % % % % % % % % % % % % % % % % % % % % % % % % % % % % % % % % % 
\subsection{Requirements}\label{sec:req}
As mentioned in the introduction, in the case of Games with a Purpose, some specific requirements emerge that motivate the need for a new truth inference approach:
\begin{description}
	\item[{[}R1{]}] \emph{Dynamic estimate of labeling quality}, by computing player reliability on control tasks: quality estimate is a usual issue in crowdsourcing, but micro-task workers may solve all the assigned tasks at once; we would like to take into account that GWAP players can play the game in different moments with different levels of attention, hence their quality/reliability can change over time and cannot be computed once and for all.
	\item[{[}R2{]}] \emph{Coping with varying difficulty of labeling task}, including possibly multiple classification or even uncertain classification tasks, which means that we cannot make any a-priori hypothesis on the number of redundant labelling actions required to solve each task.
	\item[{[}R3{]}] \emph{Incremental computation of truth inference}: as introduced in Section~\ref{sec:intro}, in GWAPs we would like to aggregate contributions as soon as they are available, because there is no pre-defined time-frame for players' input.
	\item[{[}R4{]}] \emph{Dynamic minimization of the number of required repeated labeling}, to avoid useless redundancy: if a task is ``easy'' we would like to ask fewer players to solve it, while if a task is ``hard'' we would like the task to remain longer in the game to be ``played''.
\end{description}

%%%%%%%%%%%%%%%%%%%%%%%%%%%%%%%%%%%%%%%%%%%%%%%%%%%%%%%%%%%%%%%%%%%%%%%%%%%%%%%%%%%%%%%%%%%%%%%%%%%%%%%%%%
\section{Approach description}\label{sec:approach}
	
\begin{figure}[b!]
	\centering
		\includegraphics[width=.65\textwidth]{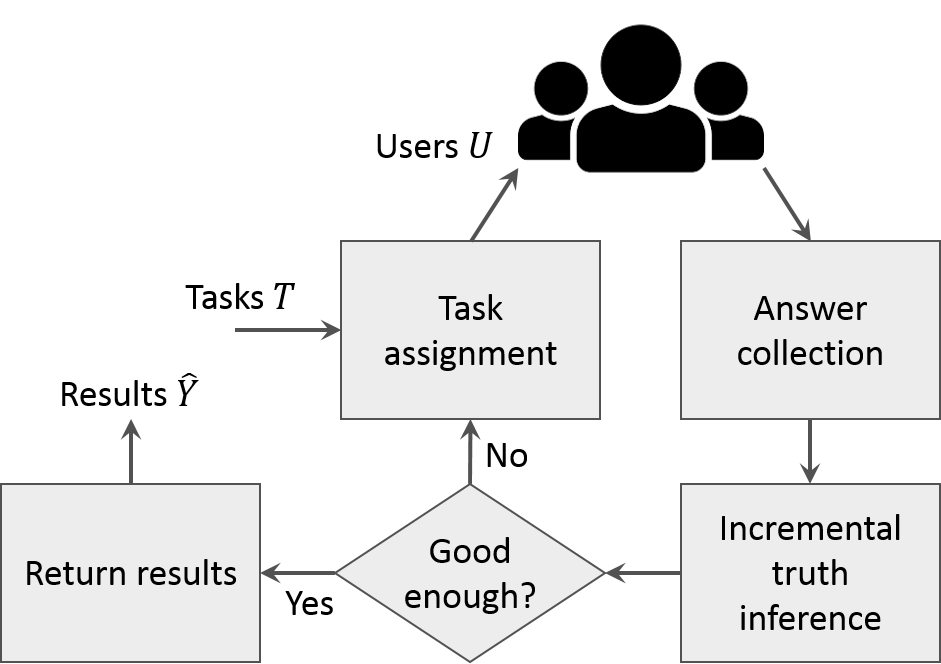}
	\caption{Incremental Truth Inference approach}
	\label{fig:approach}
\end{figure}

To address the above requirements, we define the framework illustrated in Figure~\ref{fig:approach}. Each time a player starts a game round, we assign a set of tasks to be solved, some of which are control tasks. We collect the answers from the player and we compute his/her reliability. 
Then, for each unsolved task, we perform a step of truth inference, 
and we incrementally compute a new estimation of the task solution. If the new estimation is ``good enough'' (cf. exit condition of Equation~1), the task is considered solved and removed from the game and its result returned. Otherwise, the task is kept in the game and assigned to the next user/player.

% % % % % % % % % % % % % % % % % % % % % % % % % % % % % % % % % % % % % % % % % % % % % % % % % % % % % % % % 
\subsection{Algorithm}

The approach outlined above is explained in details in the following Algorithm~\ref{algo1}. 
Each time a player starts a game round (line 2), he/she is assigned a set of tasks to be solved. 

\begin{algorithm}[htb]
	\SetKwFunction{GetActiveUser}{GetActiveUser}
	\SetKwFunction{SelectTasks}{AssignTasks}
	\SetKwFunction{SelectControlTasks}{AssignControlTasks}
	\SetKwFunction{CollectAnswer}{CrowdsourceAnswer}
	\SetKwFunction{EvaluateUserReliability}{ComputeUserReliability}
	\SetKwFunction{TrueAnswer}{TrueAnswer}
	\SetKwFunction{size}{size}
	\SetKwFunction{UpdateSolutionEstimate}{UpdateSolutionEstimate}
	\SetKwFunction{isTaskSolved}{isTaskSolved}
	\SetKwComment{Comment}{\hspace{-.5em}/*\hspace{.2em}}{\hspace{.2em}*/}
	
	\BlankLine
	\While{$T \neq \,\,$\O}{ 
		$u_k \leftarrow$ \GetActiveUser{$U$}\; 
		
		\BlankLine
		\Comment{measure user reliability on control tasks}
		$G' \leftarrow$ \SelectControlTasks{$G,u_k$}\; 
		$errors \leftarrow 0$\;
		\ForEach{$g_i$ in $G'$}{
			$c_{i,k} \leftarrow$ \CollectAnswer{$g_i,u_k$}\;
			\If{$c_{i,k} \neq$ \TrueAnswer{$g_i$}}{
				$errors \leftarrow errors+1$\;
			}
		}
		$q_k \leftarrow$ \EvaluateUserReliability{$errors,\,$\size{$G'$}}\;
		\BlankLine
		\Comment{aggregate user answers in truth inference}
		$T' \leftarrow$ \SelectTasks{$T,u_k$}\;
		\ForEach{$t_i$ in $T'$}{
			$c_{i,k} \leftarrow$ \CollectAnswer{$t_i,u_k$}\;
			\UpdateSolutionEstimate{$t_i,c_{i,k},q_k$}\;
			\If{\isTaskSolved{$t_i$}}{
				$\hat{y_i} \leftarrow c_{i,k}$\;
				$T \leftarrow T - \{t_i\}$\;
			}
		}
	}
	\Return $\hat{Y}$
	
 \caption{Incremental Truth Inference Algorithm}\label{algo1}
\end{algorithm}

The player provides answers to each task without being able to distinguish between unsolved tasks and control tasks (cf. lines 6 and 14). The answers on control tasks are used to compute player's reliability, which is a function of the number of mistakes (lines 5-11); reliability is computed per each game round. 
There are of course different ways to realize the \texttt{ComputeUserReliability} function of line 11: the simplest way is to use the percentage of correct labels in control tasks, i.e. $q_k \leftarrow 1-errors/$\texttt{size}$(G')$. In other cases, it may be safer to  strongly penalize players which submit random answers; in the games that we employ in our evaluation (cf. Section~\ref{sec:eval}), to have a conservative estimation, we adopted the following formula:
\begin{equation}\label{eq:reliab}
	q_k \leftarrow e^{-\alpha \cdot errors}
\end{equation}
where $\alpha$ is set (for example) so that $q_k$ almost halves with 1 mistake and then quickly decreases with further errors.

On the other hand, the answers on unsolved tasks are weighted with the reliability value and used to update the estimation scores (lines 14-15); for each task $t_i$ and for each possible label $v_j$, the \texttt{UpdateSolutionEstimate} function is implemented as follows:
\begin{equation}\label{eq:score}
	s_{i,j} \leftarrow \begin{cases}
		s_{i,j} + \delta \cdot q_k & \text{if } c_{i,k} = v_j, \\
		s_{i,j} & \text{otherwise}.
	\end{cases}
\end{equation}
where $c_{i,k}$ is the label contributed by the user with reliability $q_k$ and $\delta$ is an increment that depends on the minimum redundancy required for the task.

At each truth inference step, the task completion condition is checked (line 16) with Equation~\ref{eq:exit} and, if it holds, the task solution is returned and the task removed from the game (lines 17-18). 
The algorithm iterates until all tasks are solved (line 1) and truth is inferred on all tasks (line 22).

% % % % % % % % % % % % % % % % % % % % % % % % % % % % % % % % % % % % % % % % % % % % % % % % % % % % % % % % 
\subsection{Requirement satisfaction} 

Qualitatively, we now assess how the approach presented in this section addresses the requirements listed in Section~\ref{sec:req} and we discuss some of its positive consequences.

Labeling quality is controlled via the updates of the estimation scores $s_{n,l}$, incremented with players' contributions which are weighted with the reliability values $q_k$. This means that the proposed approach takes into consideration the quality of contributions and ``measures'' it at each game play, thus relying on a ``local'' trustworthiness value; the dynamic re-computation of $q_k$  fulfills requirement [R1], by addressing the fact that the same player can show a different behaviour in different moments of his/her playing, e.g. being careful vs. distracted. 

The estimation scores $s_{n,l}$, their update function (cf. Equation~\ref{eq:score}) and the task completion condition (cf. Equation~\ref{eq:exit}) have also other interesting properties. 
The scores are attributed to each task-label combination and updated at each user contribution. 

If a task $t_i$ is ``easy'', different players will attribute the same label $v_l$ and the respective score $s_{i,l}$ will quickly increase and overcome the threshold $\bar{s}$ of the exit condition. On the contrary, if a labelling task is difficult or controversial, different GWAP players may give different solutions from the set $V$ to the same task $t_i$, so potentially all scores in $s_{i,*}$ get updated but none of them easily overcomes $\bar{s}$.

In other words, the proposed approach fulfills requirements [R2] on task difficulty, because easy and difficult tasks are automatically detected and treated accordingly, and [R4] on repeated labelling, as the number of players asked to solve the same task is dynamically adjusted.

It is worth noting that in record linkage literature~\cite{fellegi1969theory}, scores are assigned to each possible couple of records, and usually the ``matching'' score is increased while the ``non-matching'' scores are decreased respectively. In the cases of possibly multiple labeling and uncertain solutions (cf. requirement [R2]), we propose to increase the score of the user-provided solution, without decreasing the score of the alternative solutions. Of course, variations of the update function in Equation~\ref{eq:score} can be introduced, depending on the scenario characteristics. For example, if $c_{i,k} \neq v_j$, then $s_{i,j}$ could be decreased of a quantity $\delta' \cdot q_k$, where $\delta'$ is the decrement amount.

By design, Algorithm~\ref{algo1} fulfills requirement [R3], since each player contribution (line 14) triggers a step of the truth inference estimate (line 15) and leads to the exit condition check (line 16). This incremental approach ensures that the task is assigned to players only until an inferred ``true'' solution is reached, thus avoiding useless redundancy of labelling (again satisfying requirement [R4]).

The dynamically adjusted repeated labelling has also the consequence of indirectly \emph{estimating task complexity}: indeed we can say that the more contributions are needed to satisfy the exit condition of Equation~\ref{eq:exit}, the more difficult the task. Therefore, whenever an assessment of the task difficulty is required, the number of collected contributions can be adopted as a proxy measure. In our previous work~\cite{re2018human} we indeed demonstrated that this empirical measure of difficulty is highly correlated with the (lack of) confidence value resulting from machine learning classifiers applied to the same data. 

A final note on task assignment: it is a common best practice to give each task to a crowd worker at most only once and to perform answer aggregation on responses from different workers; this is also true for GWAPs, in that the same player could get bored if requested to solve the same problem over and over. This means that task assignment to player $u_k$ (lines 3 and 12) takes tasks from $G$ and $T$ respectively among those that $u_k$ never solved before. A pragmatic strategy to avoid using up the entire set $G$ of control tasks, that we usually adopt when implementing GWAPs, is to dynamically increment $G$ by adding the solved tasks from the set $T$ (those removed when the ``true'' solution is inferred), so line 18 could become: $T \leftarrow T - \{t_i\};~ G \leftarrow G + \{t_i\}$.

%%%%%%%%%%%%%%%%%%%%%%%%%%%%%%%%%%%%%%%%%%%%%%%%%%%%%%%%%%%%%%%%%%%%%%%%%%%%%%%%%%%%%%
\section{Evaluation}\label{sec:eval}

To evaluate the proposed truth inference algorithm we performed a comparative assessment with alternative solutions, on the basis of the data collected through two different GWAPs: the LCV Game~\cite{brovelli2018crowdsourcing} and Night Knights~\cite{re2018human}. 

The Land Cover Validation (LCV) Game\footnote{Cf. \url{http://landcover.como.polimi.it/landcover/}.} addresses a multinomial classification of items with 5 different labels; domain experts required a minimum of 3 different and agreeing contributions for each item classification. Night Knights\footnote{Cf. \url{https://www.nightknights.eu/}.} asks players to classify pictures with one of 6 admissible labels; at least 4 agreeing contribution from different users were requested by experts on the basis of domain-specific considerations.

A first evaluation of our approach is based on the total number of contributions to be collected (in line with requirement [R4]). In most crowdsourcing settings, where aggregation is computed ex-post, a fixed number of contributions is collected per each task. Let's consider the multinomial classification of $N$ tasks with $L$ admissible labels, with a minimum of $p$ agreeing labels per task. To implement an ex-post aggregation with simple majority voting, the total number of needed contributions is the \emph{redundancy} $r$ computed as
\begin{equation}
	r \leftarrow N \cdot (\,(p-1) \cdot L + 1\,)
\label{eq:apriori}
\end{equation}
Moreover, in traditional micro-work/crowdsourcing settings, there is experimental evidence of 40-45\% of spammers among crowd workers~\cite{shah2010spam,vuurens2011much}, thus redundancy could be even higher than the one computed in Equation~\ref{eq:apriori}. 

Table~\ref{tab:risparmio-algo} shows the theoretical and empirical numbers for LCV Game and Night Knights: the incremental approach that we propose leads to a sensible ``saving'' in terms of redundancy, since whenever the minimum number $p$ of contribution is enough to consider the task solved, no more labels are sought.

\begin{table}[t]
\centering
\begin{tabular}{lcccccc}
		\toprule
		\textbf{GWAP} & \textbf{$N$} & \textbf{$L$} & \textbf{$p$} & \textbf{\begin{tabular}[c]{@{}c@{}}~Theoretical~ \\ $r$\end{tabular}} & \textbf{\begin{tabular}[c]{@{}c@{}}Actual \\ $r$\end{tabular}} & \textbf{~\% diff.} \\ 
		\midrule
		LCV Game   & \smalltilde1,000   & 5  & 3  & \smalltilde11,500    & \smalltilde6,400    & -44\%  \\
		Night Knights~   & ~\smalltilde27,700~  & ~6~  & ~4~  & \smalltilde525,000   & ~\smalltilde205,000~  & -61\%  \\
		\bottomrule \\   
\end{tabular}
	\caption{Number of required contributions for truth inference over $N$ tasks, with $L$ possible labels and a minimum of $p$ agreeing answers: comparison between theoretical redundancy $r$ (under the hypothesis of ex-post aggregation with simple majority voting) and actual numbers as experimentally measured in the two considered GWAPs applying our incremental truth inference approach. }
	\label{tab:risparmio-algo}
\end{table}

Finally, to assess the ability of our incremental approach to infer the truth, we applied state-of-the-art algorithms for ex-post data aggregation and compared the resulting classification on the contribution collected by our GWAP. Namely, we run %simple majority voting, 
expectation maximization~\cite{dawid1979maximum} and message passing~\cite{karger2011iterative}, which are the most frequently used truth inference algorithms; then, we compared the aggregated labels with a confusion matrix. The results reported in Table~\ref{tab:cfr-algo} show that indeed the overlap between the ``truths'' inferred with the compared algorithms is very high and the agreement statistics confirm it. This proves the validity and applicability of our approach.

\begin{table}[b]
\centering 
\begin{tabular}{lccccc}
		\toprule
		\textbf{GWAP}           & ~\textbf{Algorithm}~ & ~\textbf{\% diff.}~ & ~\textbf{Accuracy}~ & ~\textbf{Kappa}~ & ~\textbf{Rand} \\ \midrule
		\multirow{2}{*}{LCV Game} & EM                 & 3.9\%            & 96.1\%            & 93.4\%         & 88.7\%        \\
														& MP                 & 3.1\%            & 96.9\%            & 94.7\%         & 90.6\%        \\ \midrule
		\multirow{2}{*}{Night Knights} & EM                 & 0.3\%            & 99.7\%            & 99.4\%         & 99.4\%        \\
														& MP                 & 0.2\%            & 99.8\%            & 99.6\%         & 99.6\%        \\ \bottomrule \\
\end{tabular}
	\caption{Truth inference results comparison between our incremental approach and state of the art techniques (EM: expectation maximization, MP: message passing) along various metrics: \% of different classifications, accuracy of the confusion matrix, Kappa statistics, adjusted Rand index corrected-for-chance~\cite{rand}.}
	\label{tab:cfr-algo}
\end{table}

%%%%%%%%%%%%%%%%%%%%%%%%%%%%%%%%%%%%%%%%%%%%%%%%%%%%%%%%%%%%%%%%%%%%%%%%%%%%%%%%%%%%%%%%%%%%%%%%%%%%%%%%%%
\section{Conclusions}\label{sec:concl}

In this paper, we proposed an incremental algorithm for truth inference that satisfies the requirements emerging from the aggregation of  player contributions in Games with a Purpose. We explained and described our approach in details, highlighting the practical consequences and advantages, including the avoidance of useless redundancy with the minimization of required task solutions, and the dynamic estimation of player reliability, label quality and task difficulty.  

We also presented a comparative evaluation of the presented approach on actual data collected through two different GWAP applications, which proves the applicability and advantages of the proposed incremental truth inference.

It is worth noting that we also released as open source the GWAP Enabler~\cite{re2018framework}, a software framework to build Games with a Purpose which implements the incremental truth inference approach outlined in this paper. The two games mentioned in the evaluation section were developed on top of this framework. The interested reader can find on GitHub both the software framework\footnote{Cf. \url{https://github.com/STARS4ALL/gwap-enabler}.} and a tutorial explaining how to use and configure it\footnote{Cf. \url{https://github.com/STARS4ALL/gwap-enabler-tutorial}.}.

%%%%%%%%%%%%%%%%%%%%%%%%%%%%%%%%%%%%%%%%%%%%%%%%%%%%%%%%%%%%%%%%%%%%%%%%%%%%%%%%%%%%%%%%%%%%%%%%%%%%%%%%%%
\subsection*{Acknowledgments}
This work is partially supported by the STARS4ALL project (H2020-688135), co-funded by the European Commission. We thank Andrea Fiano for the implementation of the Night Knights game, Alejandro Sánchez de Miguel and Lucía García for their support in the interpretation of our experimental results from a light pollution research point of view and Esteban González Guardia for the retrieval and provision of images from the NASA repository.

%%%%%%%%%%%%%%%%%%%%%%%%%%%%%%%%%%%%%%%%%%%%%%%%%%%%%%%%%%%%%%%%%%%%%%%%%%%%%%%%%%%%%%%%%%%%%%%%%%%%%%%%%%
\bibliographystyle{splncs}
\bibliography{biblio} 

\end{document}